# VALUE OF USAGE AND SELLERS' LISTING BEHAVIOR IN INTERNET AUCTIONS

Sangin PARK

Department of Economics
SUNY at Stony Brook
Stony Brook, NY 11794-4384
Tel: 631-632-7559
Fax: 631-632-7516
Email: sanpark@notes.cc.sunysb.edu
Web: http://ms.cc.sunysb.edu/~sanpark

September 2001
(Very Preliminary)

In this paper, we aim to empirically examine the value of website usage and sellers' listing behavior in the two leading Internet auctions sites, eBay and Yahoo!Auctions. The descriptive data analysis of the seller's equilibrium listing behavior indicates that a seller's higher expected auction revenue from eBay is correlated with a lager number of potential bidders measured by website usage per listing. Our estimation results, based on the logarithm specifications of sellers' expected auction revenues and potential bidders' website usage, show that in a median case, (i) 1 percent increase of the unique visitors (page views) per listed item induces 0.022 (0.007) percent increase of a seller's expected auction revenue; and (ii) 1 percent increase of sellers' listings induces 1.99 (4.74) percent increase of the unique visitors (page views). Since increased expected auction revenues will induce more listings, we can infer positive feedback effects between the number of listings and website usage. Consequently, Yahoo!Auctions, which has substantially less listings, has greater incentives to increase listings via these feedback effects which are reflected in its fee schedules.

I. Introduction

For the last two and a half years, the fluctuation of the market value of an e-commerce firm has been huge. First, the potential growth of e-commerce has hit the stock market: due to the Internet related stocks, the Nasdaq composite index went up more than 70 percent in 1999. Then, the enthusiasm of the Wall Street was struck by disappointing profits and worries about slowing economy. Between December of 1999 and August of 2001, the stock price of eBay peaked at $127.50, plunged to $26.75, and then bounced back to the range of $50 to $70. During the same time period, the stock prices of Yahoo! and Amazon.com continuously plunged from the level of $250 to the level of $10 and from the level of $110 to the level of $8, respectively. The valuation of Internet stocks is particularly difficult mainly because e-commerce is still only in the beginning stage and its potential outreach is believed to be huge.[1] Hence, it is speculated that accounting information such as bottom-line net income is usually of limited use in the valuation of Internet stocks. Instead, website usage measured by 'unique visitors' or 'page views'[2] is often considered as a good proxy for the potential growth and profitability of an e-commerce firm in the future. For example, Trueman, Wong and Zhang (2000) found that Internet usage provides considerable explanatory power for the prices of some Internet stocks (before the plunge in 2000). The recent plunge of e-commerce firms' stock prices, however, may cast doubts on this argument.

In this paper, we aim to empirically examine the value of website usage in a particular e-commerce called Internet auctions. We are specifically interested in: (i) how website usage affects a seller's expected auction revenue; and (ii) whether there exist positive feedback effects between the number of listings and website usage. We begin by raising a question why website usage is a valuable (or relevant) information in Internet auctions. As indicated in the literature of auction theory (see Levin and Smith (1994); Bulow and Klemperer (1996)), the seller's expected auction revenue depends on the number of the potential bidders. Hence, when a seller decides which Internet auctions site he/she will list his/her item on, he/she may refer to website usage (as well as number of other listings) for the number of potential bidders in an Internet auctions site. In the paper, therefore, we will understand the value of website usage as a relevant information on the size of potential bidders. Specifically, we will assume that a seller's expectation for the number of potential bidders in an Internet auction site is proportional to the website usage per listed item on the site. We will also consider that website usage may be not exogenously determined but influenced by the number of listings. A potential bidder may prefer to log on to an Internet auctions site with a larger number of listings because the probability that he/she may find an item to bid for will be greater. On the other hand, in the case of the common value auctions, the bidder's expected profits may fall with additional bidders due to the effects of the winner's curse.[3] Hence, it will be an empirical question whether increased listings will induce more or less usage in Internet auctions.

---

[1] Although the potential outreach of e-commerce is believed to be huge, online sales still remain a small portion of total retail sales. In 1998, electronic retailing sales accounted for 0.5 percent of all sales, and in 1999, they are projected to account for 1.2 percent (see "Survey Show Online Sales up 300%," *The New York Times*, December 29, 1999).
[2] 'Unique visitors' is the estimated number of different individuals who visit a firm's website, and 'page views' is the number of unique visitors multiplied by the average unique pages viewed per visitor.

Based on our unique weekly data of the first 17 weeks of the year 2001, we will empirically analyze the relationship between sellers' listing behavior and website usage in the two dominant Internet auctions sites, eBay and Yahoo!Auctions, which are believed to occupy more than 90 percent of Internet auctions.[4] Indeed, the comparison between eBay and Yahoo!Auctions may be ideal for our quest for positing a simplest model for the relationship between sellers' listing behavior and website usage. In equilibrium, a seller's expected revenues from listing an item on any auction website must be the same. The seller's expected revenue from listing the item on an Internet auctions site depends on both the expected auction revenue and the fees charged by the site. The comparison of listing fees between eBay and Yahoo!Auctions indicates that a seller will expect less revenues from auctioning the same item in Yahoo!Auctions than in eBay. Since there is almost no difference in available auction formats and other characteristics between the two largest Internet auctions, differences in the seller's expected auction revenues between these two Internet auctions might be caused by different sizes of potential bidders. Our data indicate that eBay has a lager number of potential bidders measured by website usage per listed item. Hence, we can infer that a seller's expected auction revenue increases with the number of potential bidders in Internet auctions. Our data also show larger website usage and more listings on eBay.

---

[3] Bajari and Hortacsu (2001) showed that for a representative sample of eBay's coin auctions, a bidder's expected profits fall by 3.2 percent when the expected number of bidders increases by one.

To quantify the extent to which the seller's expected auction revenue is affected by the number of potential bidders and the extent to which the website usage of an Internet auctions site is affected by its own and its rival's listings, we will further specify a seller's expected auction revenue function and the website usage function. Due to virtual no difference in available auction formats and other characteristics between the two largest Internet auctions, we will, without modeling specific auction mechanism, assume a simple logarithm specification of sellers' expected auction revenues. We will also specify the website usage of an Internet auctions site as a logarithm function of the number of listings on its own site and its rival's. Our estimation results show that 1 percent increase of the unique visitors (page views) per listed item induces 0.022 (0.007) percent increase of a seller's expected auction revenue. Furthermore, 1 percent increase of sellers' listings induces 1.99 (4.74) percent increase of the unique visitors (page views). Since increased expected auction revenues will induce more listings, we can infer positive feedback effects between the number of listings and website usage. Consequently, Yahoo!Auctions, which has substantially less listings, has greater incentives to increase listings via these feedback effects which are reflected in its fee schedules.

---

[4] As quoted in Lucking-Reiley (2000), an eBay Vice President said in January 2000 that eBay's market share in the Internet auctions market has remained at approximately 90 percent, which is consistent with the

The organization of the paper is as follows. Section II provides a brief review of Internet auctions. Section III discusses sellers' equilibrium listing behavior and usage of Internet auctions sites. Section IV adopts logarithm specifications of a seller's expected auction revenue and the usage equation and discusses our quantitative analyses. Section V concludes the paper.

## II. Internet Auctions and Descriptive Statistics

An Internet auctions site, such as eBay and Yahoo!Auctions, acts as a listing agent, allowing individual sellers to register their items for its website and running Web-based automatic auctions on their behalf.[5] Actual exchanges including payment and shipment are worked out by the buyer and the seller on their own. The English auctions have been the most dominant format in Internet auctions. However, sellers usually have some control over these Web-based auctions, choosing a set of different parameters for each auction such as the duration days, an opening value, and an optional secrete reserve price. A variety of goods are auctioned in Internet auctions, but the largest category by far has been the collectibles. Each Internet auctions site has different categories, and there is usually no one top-level category that includes all the types of collectibles. During the first 17 weeks of the year 2001, 55.8 percent of listings on eBay belong to one of the categories such as 'antiques & art', 'collectibles', 'books, movies, music', 'coins & stamps', 'dolls & doll houses', or 'toys, bean bag plush'. During the same time period, 61.9 percent of listings of Yahoo!Auctions were included in one of the following categories, 'antique, art & collectibles', 'sports cards & memorabilia', 'toys & games & hobbies', or 'coins, paper money & stamps'.

---

revenue estimates in Lucking-Reiley (2000). Refer to section II for details.
[5] There have been retail merchants such as Onsale and Egghead who use the auction format to sell their products. In this paper, we narrowly define Internet auctions as consumer-to-consumer transactions via listing agents.

Internet auctions began in 1995 and have been growing rapidly. As of fall 1999, Internet auctions sites are estimated to have almost $100 million revenues per month. Since the beginning of Internet auctions, eBay has maintained a dominant leading position although the popular attention and the profitability of eBay induced entries of two biggest e-commerce firms, Yahoo! in October 1998 and Amazon.com in March 1999. As quoted in Lucking-Reiley (2000), an eBay Vice President said in January 2000 that eBay's market share in the Internet auctions market has remained at approximately 90 percent, which is consistent with the revenue estimates in Lucking-Reiley (2000). As of summer 1999, eBay had 340,000 auctions closing per day while Yahoo!Auctions and Amazon.com had only 88,000 auctions and 10,000 auctions, respectively. Refer to Lucking-Reiley (2000) for detailed surveys on Internet auctions.

Our more up-to-dated information on listings and website usage is supportive of the dominance of eBay and the growth of Yahoo!Auctions. During the first 17 weeks of the year 2001, we have countered the number of listings on both eBay and Yahoo!Auctions every Wednesday. Like most Internet auctions sites, eBay and Yahoo!Auctions allow sellers to choose their own auction length. At eBay, sellers can choose a length of 3, 5, or 7 days and a length of 10 days with an extra fee of $0.10 while at Yahoo!Auctions, sellers can choose a length between 2 and 14 days. Due to the longer possible duration days, our weekly counting of listings may relatively overstate the number of new listings on Yahoo!Auctions. However, in the survey of Lucking-Reiley (2000), a modal length of duration is 7 days.

The data of weekly website usage (unique visitors and page views) of these two auctions sites are obtained from Nielsen//NetRatings during the same time period. However, we do not have the usage information on eBay in the first week of March nor the usage information on Yahoo!Auctions in the first week of January. As illustrated in figure 1,[6] during the first 17 weeks of 2001, eBay and Yahoo!Auctions had about 5,822,000 and 3,349,000 listed items on weekly average, respectively, while Amazon.com, the third largest Internet Auctions site, had only 769,000 listed items on weekly average.[7] The dominance of eBay in Internet auctions is more obvious in terms of website usage. As illustrated in figures 2 and 3, during the same time period, eBay had about 6,250,000 unique visitors and 763,6378,000 page views on weekly average while Yahoo!Auctions had 527,000 unique visitors and 1,726,000 page views on weekly average.

Relatively small differences in the numbers of listings between eBay and Yahoo!Auctions (compared to differences in website usage and closing auctions) may be mainly due to different listing fees. eBay charges two types of basic fees to sellers: insertion fees and final value fees.[8] The insertion fees of eBay range from $0.30 to $3.30, depending on the opening values (called also reserve prices or minimum bid levels) while the final value fees are 5 percent of the sale price (called also closing value) up to $24.99, 2.5 percent from $25.00 up to $1000.00, and 1.25 percent over $1000.00. On the other hand, Yahoo!Auctions charges only insertion fees ranging from $0.20 to $1.50 (see table 1). Table 1 indicates that eBay charges slightly higher insertion fees for all the ranges of opening values. A seller can *ex ante* choose not only an opening value but also a secret reserve price.[9] The fees for the secret reserve price auctions are fully refundable if the item is sold. The basic fees of Internet auctions have not changed frequently. Indeed, Yahoo!Auctions began to charge insertion fees in the beginning of the year 2001. At the same time, eBay raised its insertion fees a little bit to the levels shown in table 1. It is noteworthy that before and after the changes of insertion fees, the number of listings on eBay increased slightly while that on Yahoo!Auctions declined considerably. As of fall 2000, eBay and Yahoo!Auctions had about 5,671,000 and 4,045,000 listed items on weekly average, respectively.

---

[6] Figure 1 may give a false impression that the number of listings has a downward trend. A longer-period observations of weekly data since February of 2000, however, indicate that the number of listings may fluctuate but have a slightly upward trend on both Internet auctions sites.
[7] Except these three largest generalist auctioneers, other small Internet auctions sites usually serve small niche markets.
[8] eBay and Yahoo!Auctions also charge non-refundable fees for several optional seller features, such as home page featured, highlight, bold, etc., which may promote the seller's listing to receive more bids.

It is also noteworthy that Amazon.com has charged $0.10 insertion fee and final value fees at the rates of eBay since it entered. Consequently, Amazon.com has had less than a quarter of the listings on Yahoo!Auctions.

III. The Model

A seller will decide which Internet auctions site he/she will list his/her item on. The seller's expected revenue from listing the item on Internet auctions site *j* depends on the expected revenue from auctioning the item and the fees charged by site *j*. There are two types of fees to sellers: insertion fees, say $F_j$, and final value fees, say $\alpha_j$. Let $R_j$ denote the seller's expected auction revenue from site *j*. Then the seller's expected revenue from listing his/her item on Internet auctions site *j* is:

(1) $\quad (1-\alpha_j)R_j - F_j$.

In equilibrium, the seller's expected revenues from listing the item on any auction website must be the same. Then for any two auctions sites, say *e* and *y*, we have in equilibrium:

(2) $\quad \dfrac{1-\alpha_e}{1-\alpha_y} = \dfrac{R_y}{R_e} + \dfrac{F_e - F_y}{R_e(1-\alpha_y)}$.

As discussed in section II, both eBay and Yahoo!Auctions charge insertion fees, ranging from $0.30 to $3.30 and $0.20 from to $1.50, respectively (see table 1). Yahoo!Auctions does not charge final value fees, but at eBay, the final value fees are 5 percent of the sale price up to

---

[9] If the secret reserve price is not met by the close of the auction, the item will not be sold. In practice, a seller sets a low opening value with a high secret reserve price to attract bidders who may drive up the

$24.99, 2.5 percent from $25.00 up to $1000.00, and 1.25 percent over $1000.00. A seller's choice of an opening value (and thus an insertion fee) may depend on the number of potential bidders. As documented in Lucking-Reiley (1999) and Bajari and Hortacsu (2001), opening value is believed to be the most important determinant of entry of bidders in Internet auctions. In practice, sellers usually set a low opening value to attract more bidders in Internet auctions.[10] The survey of Lucking-Reiley (2000) indicates that as of fall 1998, most of collectibles traded on Internet auctions are relatively inexpensive with median prices well below $100. In the U.S. mint/proof coin sets auctions on eBay, as reported in Bajari and Hortacsu (2001), the average opening value was $16.28 while the average value of the traded coins is $47. Hence, in the paper, we will consider as a median case a seller listing an expected auction value of $50 on eBay with an opening value between $0.10 and $24.99. Hence, the equilibrium condition in (2) indicates that a median seller will expect about 4 percent less revenues from auctioning the same item on Yahoo!Auctions than on eBay.[11]

As pointed out in the literature of auction theory, a seller's expected revenue auctioning on site $j$ depends on several factors such as the number of potential bidders, say $N_j$, and available auction mechanisms, say $m_j$, such as available auction formats and a set of parameters which the seller can choose in each auction. That is, $R_j = R(N_j, m_j)$, where $R$ is a real function. Since both eBay and Yahoo!Auctions offer almost the same choices of auction formats and auction parameters, we can infer that a seller's different expected auction revenues from eBay and Yahoo!Auctions mainly result from the different numbers of potential bidders. As discussed in section I, we assume that a seller's expectation for the number of potential bidders is proportional to the website usage per listed item. In fact, eBay has substantially larger website usage per listed

---

price.
[10] As indicated in Lucking-Reiley (2000), the conventional wisdom seems to be that $0 opening value plus a $50 secret reserve price would be more profitable to the seller than a $50 opening value with no secret reserve price.
[11] If the final value on eBay auctions is $50 with an insertion fee of $0.55, then the seller will pay the total fee equal to $2.425 (= $25(0.05) + $25(0.025) + $0.55) on eBay auctions, while the seller will pay only an

item (as well as substantially more listed items and larger website usage as discussed in section II) than Yahoo!Auctions: during the first 17 weeks of 2001, the average weekly unique visitors per listed item are 1.07 for eBay and 0.16 for Yahoo!Auctions, and the average weekly page views per listed item are 131.2 for eBay and 0.52 for Yahoo! Auctions (see figures 4 and 5). Therefore, the equilibrium condition of the listing behavior in (2), combined with our empirical observations on listing fees, available auction mechanisms, and sizes of potential bidders, implies that a seller expects a higher auction revenue on eBay than on Yahoo!Auctions and this higher expected auction revenue is positively correlated with a lager number of potential bidders. Note that our descriptive analysis does not assume any specific auction mechanism nor private-value or common-value settings.[12]

Equation (2) describes equilibrium listings for given website usage. Website usage, however, may be not exogenously determined but influenced by the number of listings. A potential bidder may prefer to log on to an Internet auctions site with a larger number of listings because the probability that he/she may find an item to bid for will be greater. On the other hand, in the case of the common value auctions, the bidder's expected profit may fall with additional bidders due to the effects of the winner's curse. Bajari and Hortacsu (2001) argued that their empirical findings of eBay auctions were more consistent with common-value auctions than private-value auctions. Hence, it will be an empirical question whether increased listings will induce more or less usage in Internet auctions. The comparison between eBay and Yahoo!Auctions as shown in figures 1 – 3, however, indicates that larger usage is strongly correlated with more listings in Internet auctions.

---

insertion fee of $0.35 on Yahoo!Auctions. Hence the seller will pay $2.075 more fees on eBay, which is approximately 4 percent of the final value.

[12] Amazon.com, the third largest Internet auctions site, is not included in our analysis for several reasons. First, we do not have an access to website usage data of auctions on Amazon.com. We were told that Nielsen//NetRatings did not collect usage data of auctions separately from the usage data of Amazon.com as a whole because website usage of auctions on Amazon.com is not substantially large. Second, although Amazon.com is the third largest Internet auctions site, it is considerably smaller than Yahoo!Auctions, the second largest. Lastly, Amazon.com offers various different auction formats and cannot be directly compared with the two leading sites without positing specific auction mechanisms.

In general, we can describe website usage of site $j$, say $U_j$, as a function of its own listings, say $L_j$, the others' listings, say $L_{-j}$, and the other factors, say $O_j$, such as the influence of the other auctions sites and potential bidders' unobservable preferences of site $j$.

(3)     $U_j = U(L_j, L_{-j}, O_j),$

where $U$ is a real function.

## IV. Quantitative Analyses

In this section, we will specify a seller's expected auction revenue function, $R(N_j, m_j)$, and a usage equation, $U(L_j, L_{-j}, O_j)$, in order to quantify the effects of the number of potential bidders on the seller's expected revenue and the effects of listings on website usage. Due to almost no difference in available auction formats and other characteristics between eBay and Yahoo!Auctions, we will, without modeling specific auction mechanism, assume a simple logarithm specification of a seller's expected auction revenue as follows:

(4)     $R_j = a N_j^b e^{\xi_j},$

where $a$ and $b$ are parameters, and $\xi_j$ represents any exogenous factors, such as the seller's idiosyncratic beliefs or certain weekly effects on auction revenues, which are not correlated with the sizes of potential bidders. In equation (4), $a$ reflects the influences of available auction mechanisms, which are common to both eBay and Yahoo!Auctions, while $b$ measures the elasticity of the seller's expected revenue with respect to the expected number of potential bidders. As discussed in section I, we assume that $N_j = \gamma(U_j / L_j)$ where $\gamma$ is a positive real number.

Based on the seller's expected auction revenue function in (4), we will use the equilibrium condition of listing behavior in (2) to estimate $b$, the elasticity of the seller's expected revenue with respect to the expected number of potential bidders. The estimation based directly on the equilibrium condition of (2), however, causes several difficulties. First of all, the equilibrium condition in (2) leads to a complicated nonlinear estimating equation. With 15-week aggregate listings and usage data, we may not expect to obtain any significant estimates from this complicated nonlinear estimating equation. Moreover, even with disaggregate (a huge amount of) seller-level data, the parameter $a$ then must be indexed by each item listed by sellers. In order to avoid too many different parameters (different $a$'s indexed by item), we may have to posit a certain structural model of the auction mechanism. Hence, to keep our analysis simple and robust to specific auction mechanisms, we will simplify the equilibrium condition of (2), calculating an auction premium of eBay equivalent to the difference in insertion and final value fees (between eBay and Yahoo!Auctions) in the median case. As discussed in section II, in the median case, the seller's expected revenue from auctioning his/her item on eBay is about $50 with an opening value between $10.00 and $25.00. Then in this median case, the seller will expect approximately 4 percent less auction revenue on Yahoo!Auctions,[13] and hence, the equilibrium condition in (2) can be rewritten as follows:

(5) $\quad 1 - \bar{\alpha} = R_y / R_e,$

where $\bar{\alpha} = 0.04$. Note that if the seller's expected revenue from auctioning his/her item on eBay is about $100 with an opening value between $10.00 and $25.00, then $\bar{\alpha} = 0.033$.[14]

Then from (4) and (5), we can obtain the following estimating equation:

---

[13] Refer to footnote 11.
[14] If the final value on eBay auctions is $100 with an insertion fee of $0.55, then the seller will pay the total fee equal to $3.675 (= $25(0.05) + $75(0.025) + $0.55) on eBay auctions, while the seller will pay only an

$$(6) \quad -\ln(1-\bar{\alpha}) = b\{\ln(U_e/L_e) - \ln(U_y/L_y)\} + \varepsilon,$$

where $\varepsilon = \xi_e - \xi_y$. Since $\xi_j$ is not correlated with the numbers of potential bidders, we can apply an OLS estimation procedure for (6). Table 2 reports these estimation results. Our estimation results, indicate that in the median case ($\bar{\alpha} = 0.04$), 1 percent increase of unique visitors (page views) per listed item induces 0.0217 (0.007) percent increase of a seller's expected auction revenue. In the case that a seller's expected auction revenue on eBay is about $100 with an opening value between $10.00 and $25.00 ($\bar{\alpha} = 0033$), 1 percent increase of unique visitors (page views) per listed item induces 0.0178 (0.006) percent increase of a seller's expected auction revenue. The estimates of $b$ in either case are stochastically very significant. Due to the fee structures of eBay and Yahoo!Auctions, as the seller's expected auction revenue increases, $\bar{\alpha}$ will decrease. Hence table 2 also indicates that the elasticity of the seller's expected auction revenue with respect to the number of potential bidders decreases as the seller's expected auction revenue increases.

In Internet auctions, website usage may be influenced by the number of listings. As discussed in section III, larger usage is strongly correlated with more listings in Internet auctions. To quantify the effects of listings on website usage, our quest here is for a simple specification of the usage equation in (3) without specifying the probability that a seller find an item to bid for nor the auction mechanism. We will use a logarithm specification as follows:

$$(7) \quad U_j = L_j^{\beta_1} L_{-j}^{\beta_2} e^{c+\eta_j},$$

---

insertion fee of $0.35 on Yahoo!Auctions. Hence the seller will pay $3.325 more fees on eBay, which is approximately 3.3 percent of the final value.

where $\beta_1$ and $\beta_2$ are parameters, $c$ is a constant, and $\eta_j$ reflects potential bidders' unobservable preferences of site $j$, which are correlated with listings. $\beta_1$ and $\beta_2$ measure the elasticity of website usage of an Internet auctions site with respect to its own listings and its rival's listings, respectively. $c$ and $\eta_j$ represent the other factors, $O_j$, in (3), and thus $c$ reflects the common influences of the other auctions sites on these two Internet auctions sites.

Taking logarithm on both sides of (7), we can obtain the following estimating equation:

$$(8) \quad \ln(U_j) = \beta_1 \ln(L_j) + \beta_2 \ln(L_{-j}) + c + \eta_j.$$

Since $\eta_j$ is not correlated with $L_j$ and $L_{-j}$, we will apply an OLS estimation procedure. Table 3 reports the estimation results of the estimating equation (8). All the estimates, regardless of the measurements of website usage, are statistically very significant. The estimation results indicate that 1 percent increase of sellers' listings on a site induces 1.99 (4.74) percent increase of the unique visitors (page views) on its own site and 1.88 (4.72) percent decrease of the unique visitors (page views) on its rival's site. As implied by the estimation results in table 2, if 1 percent increase in the number of listings induce more than 1 percent increase of website usage, then the expected auction revenue will be raised by these increased listings. Hence, the estimation results of table 3 indicate that an increase in listings will eventually raise the seller's expected auction revenue in an Internet auctions site. Since increased expected auction revenues will induce more listings, we can infer positive feedback effects between the number of listings and website usage. Consequently, Yahoo!Auctions, which has substantially less listings, has greater incentives to increase listings via these feedback effects which are reflected in its fee schedules.

## V. Concluding Remarks

In this paper, we empirically examine the value of website usage and the relationship between usage and listings in the two leading Internet auctions sites, eBay and Yahoo!Auctions. Our quantitative analyses show that 1 percent increase of unique visitors (page views) per listed item induces 0.022 (0.007) percent increase of a seller's expected auction revenue. Furthermore, we can infer positive feedback effects between the number of listings and website usage, which are similar to the indirect network effects documented in the market for Yellow Pages as in Rysman (2000) and in the VCR cases as in Park (2001). The existence of these positive feedback effects may provide explanations for why eBay, the start-up company in Internet auctions, has been able to maintain its dominant position and why Yahoo!Auctions, a later entrant, set lower insertion fees and no final value fees.

In Internet auctions, a seller can not restrict the number of potential bidders of a certain site but can choose an auction site based on the sizes of potential bidders. Our empirical finding invites more theoretical and empirical studies on the relationship between the number of potential bidders and a seller's expected auction revenue. The theoretical predictions on this relationship depend on whether potential bidders' entry to an auction is endogenous or not. Levin and Smith (1994), based on the equilibrium analyses of auctions with endogenous entry, concluded that the expected revenue of any seller who uses her optimal mechanism decreases with the number of potential bidders in a mixed-strategy entry equilibrium.[15] Hence, if the number of potential bidders is too high, the seller can be better off *ex ante* by restricting the number of potential bidders. On the other hand, Bulow and Klemperer (1996) showed that very generally in a private-value auction and also in a wide class of common-value auctions, a simple ascending auction with no reserve price and N+1 symmetric bidders is more profitable to the seller than any realistic auction with N of these bidders. At a first glance, our empirical finding that a seller's expected auction revenue increases with the number of potential bidders in Internet auctions seems to be more consistent with the presumption in Bulow and Klemperer (1996). However, the experimental evidence of eBay auctions in Lucking-Reiley (1999) was supportive of stochastic endogenous entry in Internet auctions.

Our intuition for this discrepancy between the theoretical prediction and our empirical finding in Internet auctions is that any Internet auctions site may not have reached the sufficient number of potential bidders (or $n^*$ in Levin and Smith (1994)). As reported in Bajari and Hortacsu (2001), the average number of bidders for U.S. mint/proof coin sets was only 3 on eBay auctions. The stochastic entry reported in Lucking-Reiley (1999) and Bajari and Hortacsu (2001) may be generated not by mixed strategies but by other things happening in bidders' lives as discussed in Lucking-Reiley (1999). In reality, sellers pay more fees for featured auctions and set low opening values (usually with higher secret reserve prices) to attract more bidders in Internet auctions. As we discussed, a huge number of website usage of eBay or Yahoo!Auctions is met by a great number of sellers listing their items. The positive feedback effects between usage and listings may be a source of growing Internet auctions but keep usage per listing from rising significantly.

---

[15] A mixed-strategy entry equilibrium occurs because there are too many potential bidders.

Table 1: Fees

|  |  | eBay | Yahoo!Auctions |
|---|---|---:|---:|
| Insertion Fees |  |  |  |
|   Opening Value | $0.01-$9.99 | $0.30 | $0.20 |
|  | $10.00-$24.99 | $0.55 | $0.35 |
|  | $25.00-49.99 | $1.10 | $0.75 |
|  | $50.00-$199.99 | $2.20 | $1.50 |
|  | $200.00 and up | $3.30 | $1.50 |
| Final Value Fees |  |  |  |
|   Closing Value | $0-$25 | 5% | Free |
|  | $25-$1000 | 2.50% | Free |
|  | over $1000 | 1.25% | Free |

Table 2: Regression of Expected Auction Revenues

| Elasticity w.r.t. potential bidders | estimate | standard error |
|---|---:|---:|
| alpha bar = 0.04 | | |
|   Usage is measured by unique visitors | 0.0216 | 0.0004 |
|   Usage is measured by page views | 0.0074 | 0.00007 |
| | | |
| alpha bar = 0.033 | | |
|   Usage is measured by unique visitors | 0.0178 | 0.0003 |
|   Usage is measured by page views | 0.0061 | 0.00006 |
| | | |
| alpha bar = 0.025 | | |
|   Usage is measured by unique visitors | 0.0134 | 0.0003 |
|   Usage is measured by page views | 0.0046 | 0.00004 |
| Number of observations | 15 | |

Table 3: Regression of Website Usage

| Variables | estimate | standard error |
|---|---:|---:|
| Usage is measured by unique visitors | | |
|   constant | 6.564 | 1.875 |
|   own listings | 1.989 | 0.146 |
|   rival's listings | -1.876 | 0.146 |
| | R-squared | 0.94 |
| Usage is measured by page views | | |
|   constant | 10.289 | 4.986 |
|   own listings | 4.743 | 0.389 |
|   rival's listings | -4.718 | 0.389 |
| | R-squared | 0.93 |
| Number of observations | 30 | |

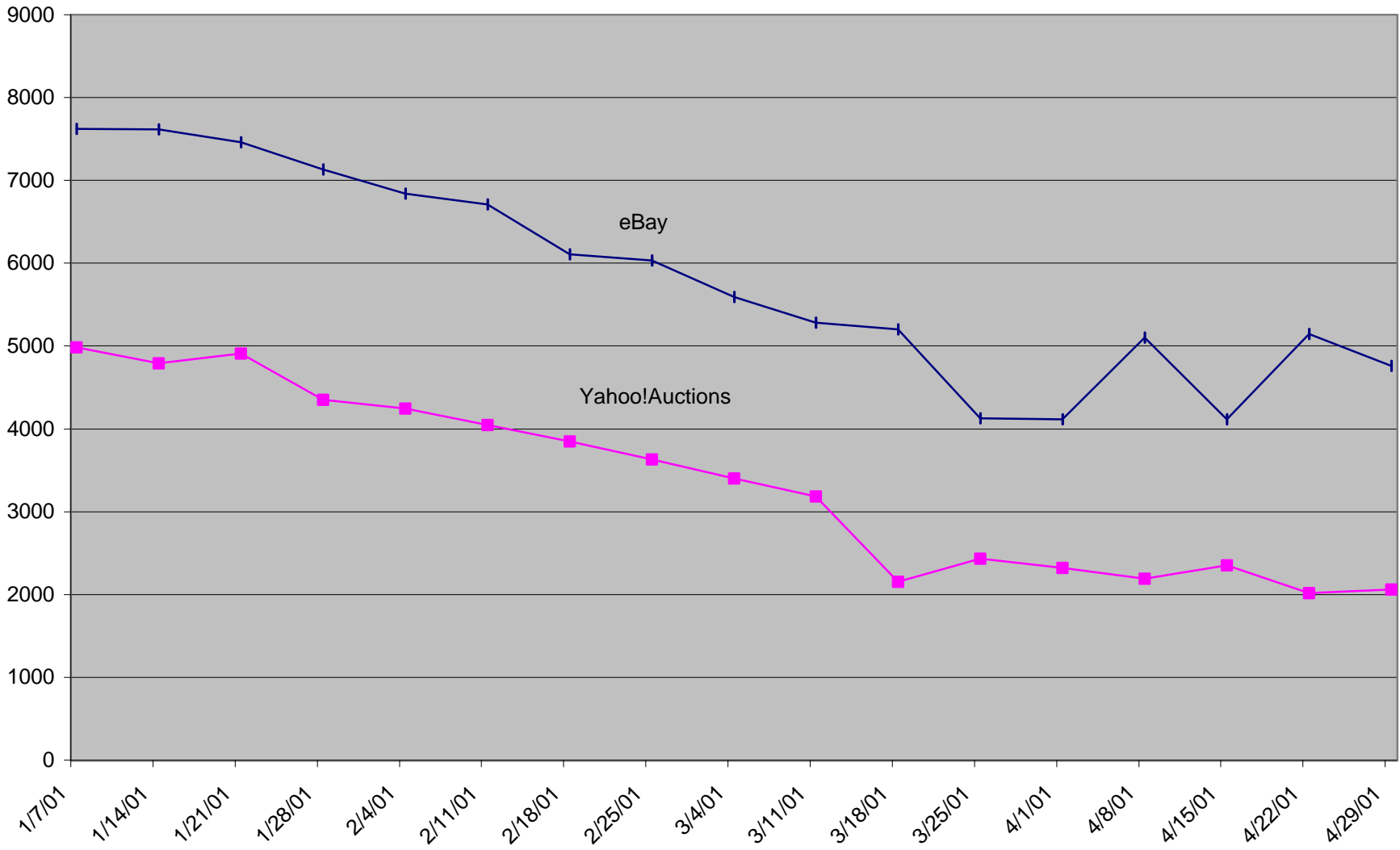

Figure 1: Listings (unit: thousands)

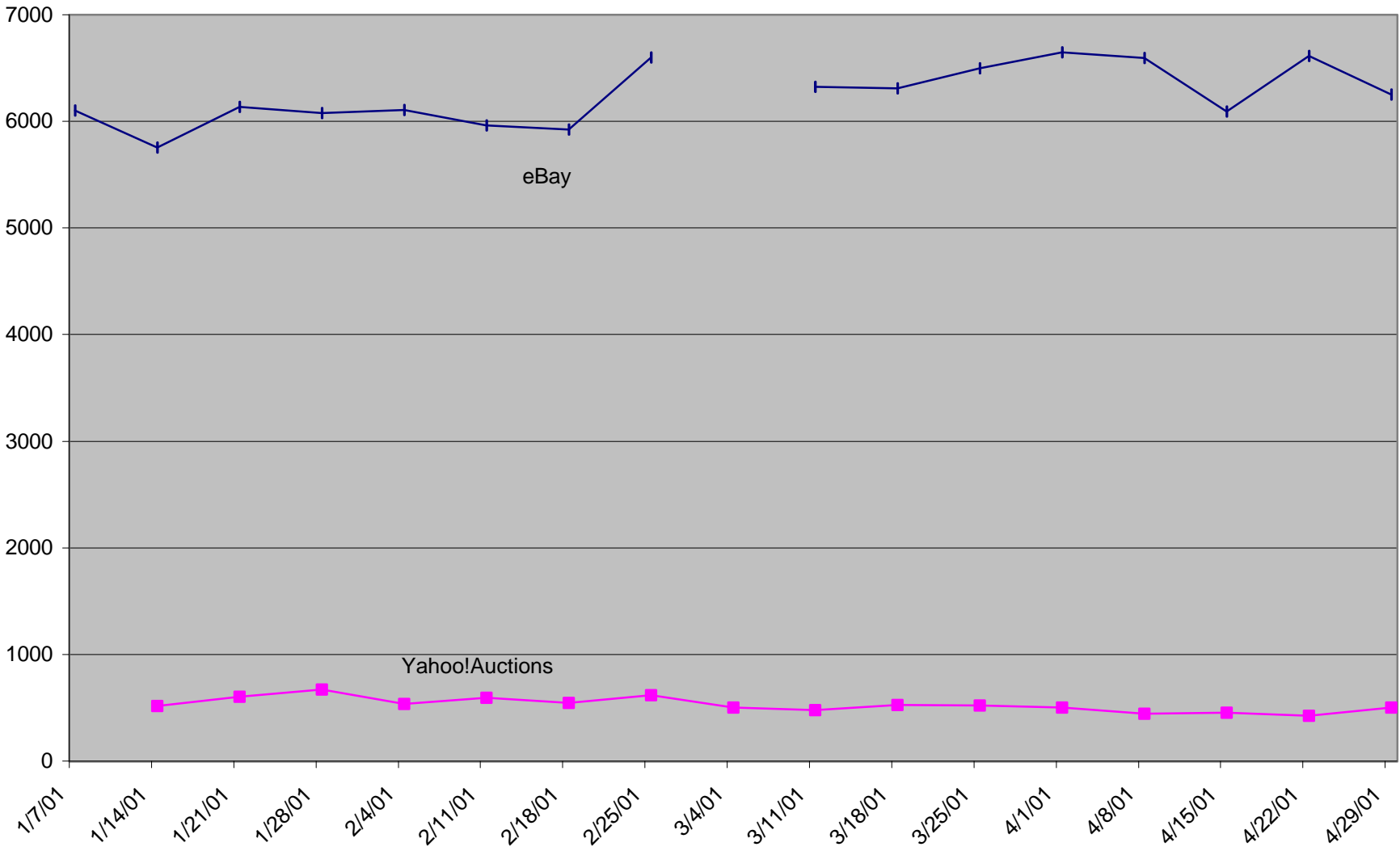

**Figure 2: Unique Visitors (unit: thousands)**

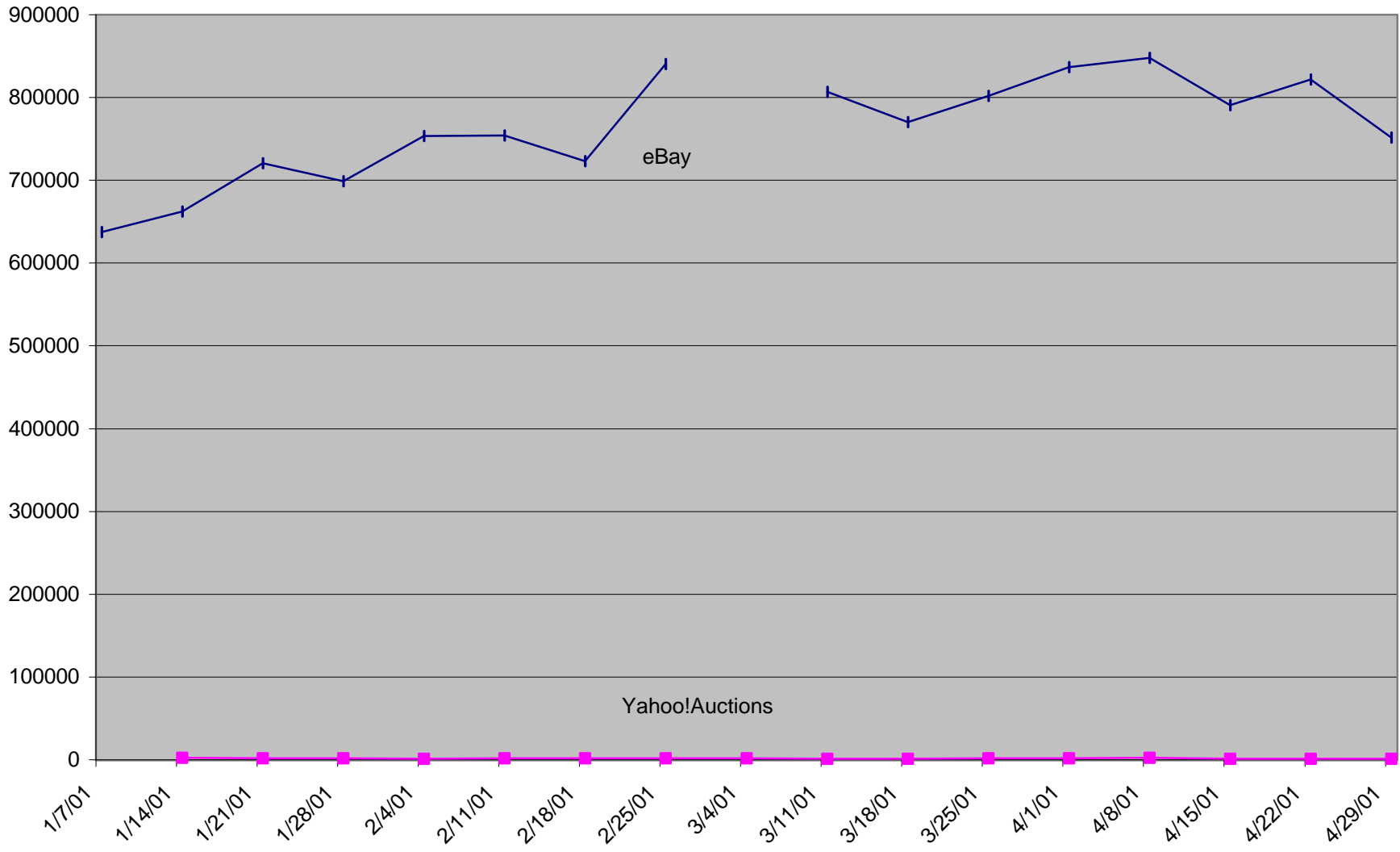

Figure 3: Page Views (unit: thousands)

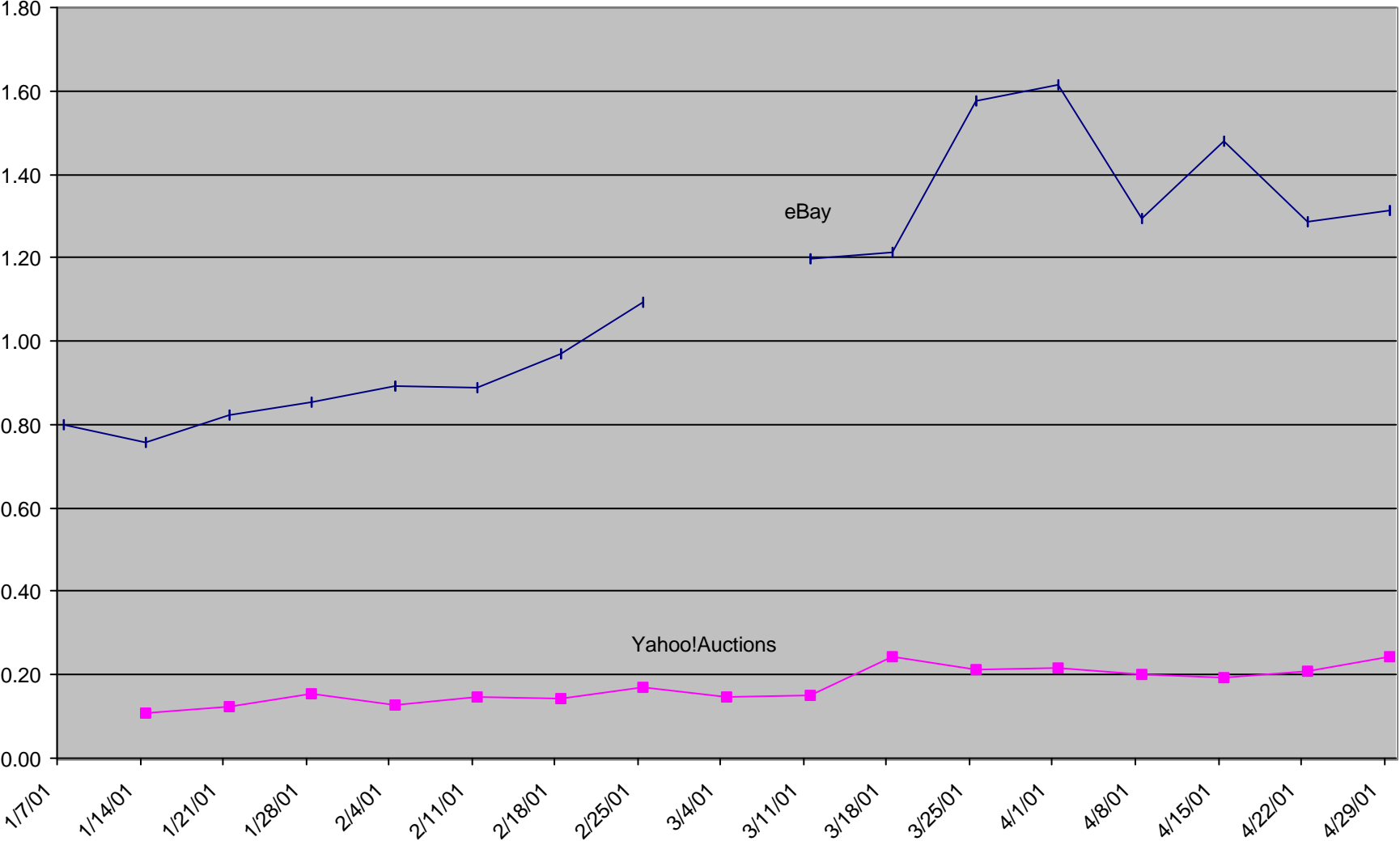

**Figure 4: Unique Visitors / Listings**

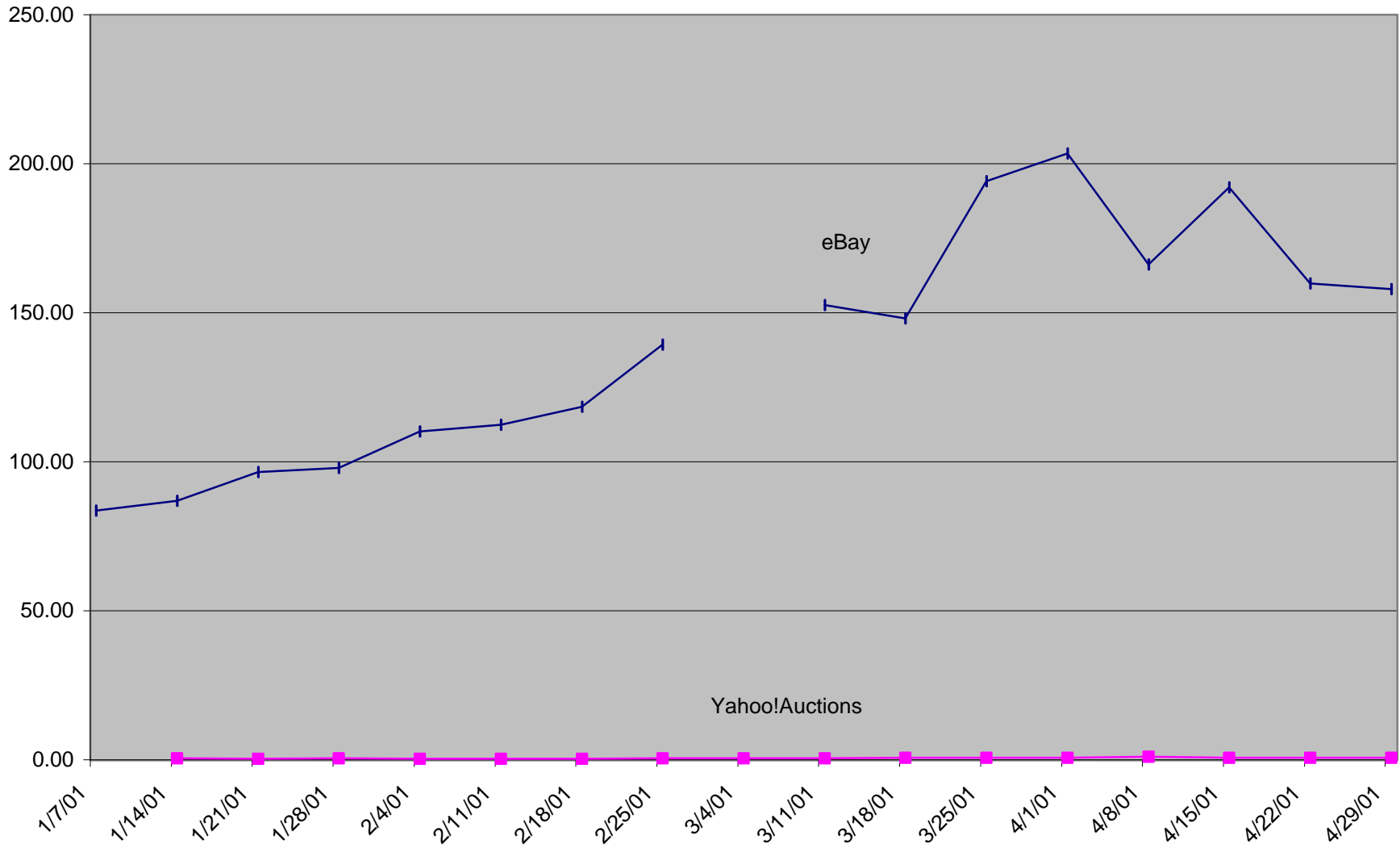

Figure 5: Page Views / Listings